\documentclass[onecolumn,amsmath,amsfonts,amssymb]{revtex4} 
\usepackage[skip=-6pt]{caption}
\usepackage{epsfig,psfrag,graphicx,bm,color}
\usepackage{graphicx}
\usepackage{mathptmx}      
\usepackage{latexsym}
\def\beq{\begin{equation}}
\def\eeq{\end{equation}}
\def\bea{\begin{eqnarray}}

\def\eea{\end{eqnarray}}

\begin{document} 

\title{ Black holes with toroidal horizons in ($d+1$)-dimensional space-time}%

\author{Elham sharifian}
\email{e.sharifian@ph.iut.ac.ir}
\author{Behrouz Mirza}
\email{b.mirza@cc.iut.ac.ir}
\author{Zahra Mirzaiyan}
\email{Z.Mirzaiyan@ph.iut.ac.ir}

\affiliation{Department of Physics,
Isfahan University of Technology, Isfahan 84156-83111, Iran}

\begin{abstract} 
We investigate black holes with toroidal horizons in ($d+1$)-dimensional space-time. Using the solution phase space method, we calculated conserved charges for these black holes before exploring some features of this metric including its entropy and thermodynamic quantities. Another aspect of the study involves obtaining a general exact static interior solution for uncharged black holes with toroidal horizons in ($d+1$)-dimensional space-time. Finally, an interior solution for charged black holes is obtained.


\end{abstract}

\maketitle

\section{Introduction}\label{a}
Study of exact solutions of the Einstein's equations is an important part of the theory of general relativity. Black hole solutions in ($2+1$)-dimensional AdS space time were first obtained by Banados, Teitelboim, and Zanelli (BTZ) \cite{BTZ}. These solutions have provided a simplified machinery for investigating different aspects of physics including gravitational interactions in lower dimensions, quantum aspects of three dimensional gravity, and the AdS/CFT correspondence \cite{witten1,witten2,maldacena,emparan,carlip1}.
In addition, the subject of black hole thermodynamics has come to be one of the most important issues in black hole's physics over the past few decades \cite{Bek1,barden,Bek2,hawking2}. The study of thermodynamic properties of AdS black holes dates back to the paper by Hawking and Page \cite{hawking1}. Since then, it has turned out that there is a rich phase structure for BHs. The thermodynamics of BTZ black holes and their stability under appropriate boundary conditions have been investigated in \cite{ghorizon,lemos,mann,haung}.

Recently, \cite{hendi} proposed an exterior solution for BTZ-like black holes in higher dimensions. In the present paper, we explore some of the properties of black holes with toroidal horizons in higher dimensions. Solution Phase Space Method (SPSM) has been recently proposed for calculating conserved charges based on the exact symmetries of solutions \cite{hajian1}. Using SPSM guarantees that the conserved charges are regular and that no divergency will appear during the process. We use this method to find the mass as the conserved charge of this solution. We will see that the parameter introduced as a constant in this metric is the ADM mass of this solution.\\
Wave equation of a non-minimally coupled scalar field has been recently investigated in the BTZ background \cite{gray}. Drawing upon previous works, we not only attempt to find a Schrodinger-like wave equation and an expression for the related potential in higher dimensions for the black holes with toroidal horizons (BHTH) but also try to find some of the thermodynamic quantities for them in arbitrary dimensions. These quantities are useful for investigating the thermodynamics and phase transitions of these black holes in future research.\\

Also, in the relativistic astrophysics, it is interesting to study the static anisotropic spheres. Finding the interior metric of such an object can be useful for describing the interior region of stellar objects. In \cite{faroogh1,faroogh2}, simplified
models such as interior solutions for BTZ black holes for charged and uncharged cases are suggested based on the assumption that the interior region of these black holes are filled with an anisotropic fluid. In this study, we obtain exact interior solutions for $(d+1)$-dimensional black holes with toroidal horizons. \\

This paper is organized as follows: After introducing a general form of the metric for black holes with toroidal horizons in ($d+1$)-dimensional space-time, we use the SPSM method to obtain the conserved charges of these solutions in Section II. In Section III, the entropy of these black holes is calculated using the Wald formula. Also, some of the thermodynamic quantities are calculated and the geometric-invariant quantities corresponding to these black holes are determined. In Section IV, we write the wave equation of a non-minimally coupled scalar field in a $(d+1)$-dimensional black hole background to obtain the effective potential barrier. An interior solution of the non-rotating $(d+1)$-dimensional black holes with toroidal horizons is obtained for the uncharged case in Section V and for the charged one in Section VI. Section VII is devoted to conclusions.\\


\section{Black holes with toroidal horizons in $(d+1)$-dimensions }\label{S1}
Einstein equations in the ($2+1$)-dimensional space-time with a negative cosmological constant admit a black hole solution \cite{BTZ}.
The action of the (2+1)-dimensional Einstein gravity in the presence of a cosmological constant is given by (we set $8G=c=1$):
\begin{equation}\label{action}
I=\frac{1}{2\pi }\int{\sqrt{-g}(R-2\Lambda ){{d}^{3}x}+B},
\end{equation}
where, $B$ is a surface term and $\Lambda $ is the negative cosmological constant. The equations of motion derived from equation (\ref{action}) are solved to yield the following form of the metric \cite{BTZ}:

\begin{eqnarray}\label{metric2+1}
d{{s}^{2}}=-N^{2}(r)d{{t}^{2}}+\frac{d{{r}^{2}}}{{N^{2}(r)}}+{{r}^{2}}{d\phi}^{2},
\end{eqnarray}
where, the lapse function ${{N}^{2}}(r)$ is given by:
\begin{equation}\label{N2}
{{N}^{2}}(r)=-m-{\Lambda} {r}^{2},
\end{equation}

\noindent with $-\infty< t<\infty $ , $0< r<\infty $ and $0\le \phi \le 2\pi $ and $m$ is a parameter related to the mass of the black hole.
Our goal is to obtain the general form of the exterior metric for ($d+1$)-dimensional black holes with toroidal horizons. We begin by considering the general form of the metric around a non-rotating black hole that has the isometry $SO(2)^{d-1}$ or Torus symmetry ($T^{d-1}$) as:

\begin{equation}\label{metricd}
d{{s}^{2}}=-N^{2}(r) d{{t}^{2}}+\frac{d r^{2}}{N^{2}(r)}+{{r}^{2}}\sum\limits_{i=1}^{d-1}{d{\phi} _{i}^{2}},
\end{equation}

\noindent where, ${\phi }_{i}$ represents the periodic coordinates. The variation with respect to the metric yields the following Einstein equations with a non-zero cosmological constant
\begin{equation}\label{ricci}
{{R}_{\mu \nu }}=\frac{1}{2}{{g}_{\mu \nu }}(R-2\Lambda ).
\end{equation}
Calculations using (\ref{metricd}) and (\ref{ricci}) yield the general form of the differential equation in ($d+1$)-dimensions for the lapse function as follows:
\begin{eqnarray}\label{Diffeq}
(d-1)N{N}''{{r}^{2}}+3{{(d-1)}^{2}}N{N}'r+(d-1){{{N}'}^{2}}{{r}^{2}}\nonumber\\
+2d \ \Lambda {{r}^{2}}+(d-2){{(d-1)}^{2}}{N^{2}}=0.
\end{eqnarray}
Solving this differential equation for $d=2$ leads to
\begin{equation}\label{N2d}
{{N}^{2}}(r)=-\Lambda {{r}^{2}}-{{C}_{1}},
\end{equation}
where, $C_{{1}}$ is the integration constants which can be determined by comparing with the BTZ solution
\begin{equation}\label{C1}
{{C}_{1}}=m.
\end{equation}

We keep this assumption in solving the differential equation (\ref{Diffeq}) for other dimensions and find the general form for lapse function in ($d+1$)-dimensions
\begin{equation}\label{glaps}
N^{2}(r)=-\frac{\Lambda {{r}^{2}}}{\frac{d(d-1)}{2}}-\frac{m}{\frac{d}{2}{{r}^{d-2}}}.
\end{equation}

Recently, a method called the "Solution Phase Space Method" has been proposed for calculating the conserved charges based on the exact symmetries of solutions \cite{hajian1}. This method is completely horizon independent since the
conserved charges can be calculated by integrating over any arbitrary 2-dimensional, spacelike hypersurface; hence, it guarantees that the conserved charges are regular and no divergency will appear. In general, the Solution Phase Space Method (SPSM) \cite{hajian1,ashtekar,lee,zoupas,kim} can be used for calculating the conserved charges of an action in gravitational theories including the mass, angular momenta, entropies, and electric charges associated with black holes in arbitrary dimensions. Using SPSM, it can be shown that the parameter $M$, used in Eq. (\ref{glaps}) is really the conserved charge of the ($d+1$)-dimensional black hole solutions. \\

The Lagrangian which we will focus on in the ($d+1$)-dimensional space-time is
\begin{equation}\label{lag}
L=\frac{1}{2 \pi} f(R)= \frac{1}{2 \pi} (R-2\Lambda),
\end{equation}
where, $R$ is the Ricci scalar and $\Lambda$ is the cosmological constant.
The equation of motion for the Lagrangian (\ref{lag}) is as follows:

\begin{equation}\label{eom}
E_{\mu\nu}=\frac{1}{2}(R-2\Lambda) g_{\mu\nu}-R_{\mu\nu}.
\end{equation}

By varying of the Lagrangian (\ref{lag}) and imposimg the equation of motion (\ref{eom}), the surface $d-2$ form will take the following form:
\begin{equation}
\Theta=\frac{\sqrt{-g}}{d!} \epsilon_{\mu \mu_{1}...\mu_{d}} {\theta}_{f}^{\mu} dx^{\mu_{1}}...dx^{\mu_{d}},
\end{equation}
where, ${\theta}_{f}^{\mu}$ is

\begin{equation}\label{teta}
{\theta}_{f}^{\mu}=\frac{1}{2 \pi} ({\nabla}_{\alpha} h^{\mu\alpha}-{\nabla}^{\mu} h).
\end{equation}

To calculate the conserved charge, $H_\eta$, associated with the exact symmetry $\eta$ of $(d+1)$-dimensional solution, one should do the following integration on the boundary of the space-like hypersurface:

\begin{equation}
{\hat{\delta}}{H_\eta}=\oint K_{\eta}(\delta{\hat{g}_{{\alpha}{\beta}}},\hat{g}_{{\alpha}{\beta}}),
\end{equation}
with,

\begin{equation}
K_{\eta}^{EH}(\delta{\hat{g}_{{\alpha}{\beta}}},\hat{g}_{{\alpha}{\beta}})=\frac{\sqrt{-g}}{2! 2!} \epsilon_{\mu\nu\rho\sigma} K_{\eta}^{ EH \mu\nu} dx^{\rho}\wedge dx^{\sigma},
\end{equation}
where,
\begin{eqnarray}\label{EH}
K_{\eta}^{EH \mu\nu} &=&\frac{1}{2\pi} (h^{\mu\alpha} {\nabla}_{\alpha} {\xi}^{\nu} -{\nabla}^{\mu} h^{\nu\alpha} -\frac{1}{2} h {\nabla}^{\mu} {\xi}^{\nu})\nonumber\\
&&-{\theta}_{f}^{\mu} {\xi}^{\nu}-[\mu\longleftrightarrow \nu],
\end{eqnarray}

\noindent and ${{\delta}{\hat{{g}}_{{\alpha}{\beta}}}}={{\frac{\partial {{\hat{g}}_{{\alpha}{\beta}}}}{\partial {m}}}{{\delta}{m}}}$ is the parametric variation of the metric. For the Einstein-Hilbert theory, we may use Eq.(\ref{teta}) to rewrite Eq.(\ref{EH}) as follows \cite{hajian2}:


\begin{eqnarray}
\label{M}
K^{\mu \nu}_{\eta}&=\frac{1}{2\pi}([\xi ^{\nu}\nabla ^{\mu}h- \xi ^{\nu}\nabla _{\tau} h^{\mu \tau} +\xi _{\tau}\nabla ^{\nu} h^{\mu \tau} + \nonumber \\
&\frac{1}{2}h\nabla ^{\nu}\xi ^{\mu}-h^{\tau \nu}\nabla _{\tau}\xi ^{\mu}]-[\mu \longleftrightarrow \nu]),
\end{eqnarray}

\noindent where, the variation of the metric tensor is defined as ${\delta}{g_{{\mu}{\nu}}}\equiv{h_{{\mu}{\nu}}}$ and $h\equiv{{h}_{\mu}}^{\mu}$ is the trace of $h_{{\mu}{\nu}}$. Also, $\xi^{\mu}$ is the associated killing vector and is determined based on the conserved charge selected.\\

To calculate the conserved charge, $H_\eta$, associated with suggested solutions in Section I, it is sufficient to perform the following integral. For a metric with no gauge field, we have $\eta=\xi$ in which $\xi={\xi^\mu}{\partial_\mu}$ refers to the symmetry with its corresponding killing vector $\xi^\mu$. If, for simplicity, surfaces of constant $(t,r)$ are chosen to integrate over, the conserved charge variation for an exact symmetry $\eta$ can be read through

\begin{equation}\label{CC}
{\hat{\delta}}{H_\eta}={\prod\limits_{i=1}^{d-1}} \int\limits_{0}^{2\pi}{{{\sqrt{-g}} \ {{K_\eta}^{tr}{{(\delta{\hat{g}_{{\alpha}{\beta}}},\hat{g}_{{\alpha}{\beta}})}}} \ {d{\phi_i}}}}.
\end{equation}

The exact symmetry for this solution is the stationary isometry which is generated by $\xi=\partial_{t}$. To find the mass as the conserved charge, we set $\eta={\partial_{t}}$.
Calculation of Eq.(\ref{CC}) results in:

\begin{equation}\label{masss}
{\hat{\delta}}{{M}}={\frac{(d-1)}{d}}{\frac{{{{(2\pi)}^{(d-1)}} }}{\pi}}{{\delta}m} \ \Rightarrow \ {M}={\frac{(d-1)}{d}}{\frac{{{{(2\pi)}^{(d-1)}} }}{\pi}}{m},
\end{equation}

Equations (\ref{masss}) shows that $M$ is the conserved charge and the mass of the ($d+1$)-dimensional black holes with toroidal horizons. The calculation procedures for the conserved charges of some other black holes may be found in \cite{hajian3}. \\

Eq. (\ref{metricd}) represents the general form of metrics for black holes with toroidal horizons in higher dimensions. This metric describes the exterior space-time of the black hole in any dimensions. Note that these solutions are not asymptoically $AdS_{d+1}$ while BTZ solutions are asymptoically $AdS_{3}$.

Since there is ($d-1$) periodic angles in ($d+1$)-dimensional space-times, the symmetry group of these space-times is the direct product of ($d-1$) of $SO(2)$ rotation groups ($SO(2)\otimes....\otimes SO(2)$). This symmetry means that the non-rotating black holes have a kind of torus-like symmetry called toroidal symmetry. The topology of the horizon is $T^{d-1}$. Also these solutions are are not locally maximally symmetric, in contrast with the BTZ black hole. Another difference of BTZ black holes with our solutions is that no gravitational waves propagate in BTZ space-time since the gravity has no degrees of freedom in 3 dimensions \cite{lora}. There are interesting proposals for generalization of three dimensional constant curvature black holes to higher dimensions somehow similar to our method while keeping the locally maximally symmetric property \cite{const1,const2,const3,local}.

It is interesting to note that solutions with toroidal horizons do not generate Closed Timelike Curves or so called CTCs. Generating CTCs is a generic feature of some solutions with a rotating cylindrical symmetry. For a review see \cite{CTC}, where one can find how a solution can have this feature.

\section{Thermodynamic and coordinate-invariant quantities of non-rotating black holes with toroidal symmetry in ($d+1$)-dimensional space-time}\label{S2}

To study the general features of the metric established in the previous section, we calculate some useful quantities.\\
Calculations show that this space-time has one horizon that can be found by solving ${{N}^{2}}(r)=0$. The Anti-de-Sitter radius, $l$, is related to the cosmological constant by $\Lambda =\frac{-d(d-1)}{2l^2}$ in ($d+1$)-dimensions.\\
The horizon is located in
\begin{equation}\label{ horizon}
{{r}_{h }}={{(\frac{ M{{l}^{2}}}{{(d-1)}(2 \pi)^{(d-2)}})}^{\frac{1}{d}}}.
\end{equation}
Now, we can rewrite relation (\ref{glaps}) in terms of ${{r}_{h}}$ and $l$ as
\begin{equation}\label{glapsr}
N^{2}(r)=\frac{(r^d -{r_{h}}^d)}{l^{2} r^{d-2}}.
\end{equation}

We then find the mass in ($d+1$) dimensions as follows:
\begin{equation}\label{mass}
M=\frac{ \ {(d-1) (2 \pi)^{(d-2)}{r}_{h}}^{d}}{{{l}^{2}}}.
\end{equation}

In order to find the line element for the $(d-1)$ dimensional horizon, the following definition is used:
\begin{equation}\label{surface element}
d{{\sigma }^{d-1}}={{r}^{2}}\sum\limits_{i=1}^{d-1}{d\phi _{i}^{2}}.
\end{equation}
Then, the area of the black hole horizon is
\begin{equation}\label{surface}
{{A}_{H}}=\prod\limits_{i=1}^{d-1}{\int_{0}^{2\pi}}{{d\phi}_{i}}{\sqrt{\gamma}}={{({2\pi}{{r}_{h}})}^{(d-1)}},
\end{equation}
where, $\gamma={({{2\pi}{{r}_{h}}})^{(d-1)}} $ is the determinant of the metric on the black hole horizon.
The entropy of the black hole can be obtained using the Wald formula \cite{gilani,Wald}:
\begin{equation}\label{entropy}
{{S}_{bh}}\equiv -2\pi \int{{{d}^{d-1}}x\sqrt{\gamma }\frac{\partial L}{\partial {{R}_{abcd}}}{{\varepsilon }_{ab}}{{\varepsilon }_{cd}}},
\end{equation}
where, the action is
\begin{equation}\label{E action}
I=\frac{1}{2\pi }\int{{{d}^{d-1}}}x\sqrt{{\gamma}}(R-2\Lambda ).
\end{equation}
A simple calculation yields the following entropy:
\begin{equation}\label{Entropy 2}
S={-{2{g}^{tt}}{{g}^{rr}}}{{A}_{H}}.
\end{equation}
We see that ${{g}^{rr}}{{g}^{tt}}$ is always equal to $-1$ in the $(d - 1)$-dimensional horizon, so
\begin{equation}\label{Entropy3}
S=2{{A}_{H}},
\end{equation}
${{A}_{H}}$ is the area of the event horizon of the black hole. We find $S=2{{A}_{H}}$ because we set $8G=1$ in action (\ref{action}). We may also find the form of the metric (\ref{metricd}) using the lapse function (\ref{glaps}) based on this same assumption. If we did not assume this, we would have the conventional relation $S=\frac{{A}_{H}}{4 G}$. \\

As we know, the first law of thermodynamics for a black hole with mass $M$ is

\begin{equation}\label{first law}
{\it dM}=T{\it dS}.
\end{equation}

\noindent Also, the Hawking temperature \cite{hawking1} of the outer event horizon surface of the (d+1)-dimensional black hole is
\begin{equation}\label{tempreture}
T={{(\frac{\partial M}{\partial S})}}={\frac{d {r_h}}{4 \pi l^2}}.
\end{equation}
\noindent For $d=2$, we have the similar temperature of the BTZ black hole \cite{BTZ,carlip1} as follows:
\begin{equation}\label{temp 2d}
T=\frac{{{r_h}}}{2\pi{l^2}}.
\end{equation}

Now, it would be interesting to find some coordinate-invariant quantities. Coordinate invariants are a set of scalars which are constructed from the Riemann, Weyl, Ricci, and other tensors in general relativity. These quantities represent the geometric properties of the space-time and they also play significant roles in classifying space-times \cite{scalar,weinberg,see inside}. In Table 1, some of these quantities are given for solutions with toroidal symmetry in ($d+1$) dimensions. It should be noted that according to coordinate-invariant quantities in Table 1, $r=0$ is the singularity of the solution and it is located inside the horizon. It is also useful to investigate the Penrose diagram of our solutions. The Penrose diagram of our solutions are exactly similar to the diagram of a BTZ black hole but every point in the Penrose diagram of BTZ solution is a circle while in the Penrose diagram of our solutions every point is a $(d-1)$-Torus ($T^{d-1}$). Again from the difference in definition of Penrose diagram we can see that these solutions and BTZ solution have different properties.
\begin{table*}
\caption{Coordinate-invariant quantities}
\label{mixedensemble}
\begin{tabular*}{\textwidth}{@{\extracolsep{\fill}}lrrl@{}}
\hline
Quantity\\
\hline

$R $ \ (Ricci scalar) &$\frac{-d(d+1)}{ l^2}$ \\
$R_{\mu \nu}R^{\mu\nu}$(Quadratic Ricci tensor scalar) & $\frac{{d^2} (d+1)}{l^4}$ \\
$R_{\mu\nu\alpha\beta}R^{\mu\nu\alpha\beta}$(Kretchmann) &$\frac{2{d}{(d+1)}}{l^4}+\frac{4{(d-1)^2}{(d-2)}{m^2}}{{d} \ {{r}^{2d}}}$ \\
$w_{\mu\nu\alpha\beta}w^{\mu\nu\alpha\beta}$(Weyl scalar) &$\frac{4{(d-1)^2}{(d-2)}{m^2}}{{d} \ {{r}^{2d}}}$ \\

\hline
\end{tabular*}
\end{table*}

\section{Potential barrier for $(d+1)$ dimensional black holes with toroidal horizons}\label{S6}

Hawking radiation has always been considered as a black body (thermal radiation) with the temperature $T_{H}$ although we know that this statement is partially true because the emitted particles from the black hole feel an effective potential barrier in the external region of the event horizon.The absorption cross section, or the so-called gray body factor, measures the modification of the original black body radiation. We know that Hawking radiation has not been detected yet but the important statement here is that gray body factor modifies the spectrum in the region where we have the maximum pair production. The potential barrier and the gray body factor for BTZ black hole have been recently given in \cite{gray}. In this section, we obtain the potential barrier for $(d+1)$-dimensional solutions with toroidal horizons and investigate its dependance on both the dimension of space-time and the value of coupling constant.\\
We consider the black hole metric in $(d+1)$ dimension as
\begin{equation}\label{metricbtz2}
d{{s}^{2}}=-f(r) d{{t}^{2}}+\frac{d r^{2}}{f(r)}+{{r}^{2}}\sum\limits_{i=1}^{d-1}{d{\phi} _{i}^{2}},
\end{equation}

\begin{equation}
f(r)=\frac{r^2}{l^2}-\frac{m}{\frac{d}{2} r^{d-2}}.
\end{equation}
where, $m$ is related to the physical mass of the black hole. In this case, we have a scalar field with the non-zero coupling $\xi$ coupled to the scalar curvature in the ($d+1$) dimensional black hole background in Eq.(\ref{metricbtz2}), described by the action

\begin{equation}\label{grayaction}
S=\frac{1}{2} \int d^{d+1} x [(\partial {\Phi})^{2}+\xi R_{d+1} {\Phi}^{2}].
\end{equation}

The wave equation of the scalar field reads as follows
\begin{equation}\label{eqs}
\frac{1}{\sqrt{-d}} {\partial}_{\mu} (\sqrt{-g} g^{\mu\nu} {\partial}_{\nu}){\Phi}=\xi R_{d+1} {\Phi},
\end{equation}

\noindent where, the non-minimal coupling $\xi$ is positive and $R_{d+1}=\frac{-d(d+1)}{l^{2}}$ is the Ricci scalar in the ($d+1$) dimensions background. Assuming the following ansatz for the wave function

\begin{equation}\label{eqs33}
{\Phi} (t,r,{\phi}_{1},.....,{\phi}_{d-1})=R(r) e^{-i\omega t} e^{i\sum\limits_{i=1}^{d-1} {m}_{i} {\phi _{i}}},
\end{equation}
\noindent we can simplify Eq.(\ref{eqs}). To simplify our calculation we assume that $m_{1}=.....=m_{d-1}=\widetilde{ m}$. It should be noted that this choice would pick a small subset of generic linearized perturbations for the scalar field. We can obtain the following differential radial wave equation in $(d+1)$ dimensions:
\begin{eqnarray}\label{diffe}
&& R^{\prime\prime}(r)+(\frac{(d-1)}{r}+\frac{f^{\prime}}{f})R^{\prime}(r)\nonumber\\
&&+(\frac{\omega^{2}}{f^2}-\frac{(d-1)}{r^2} \frac{m^2}{f}-\frac{(d-1)\xi R_{d+1}}{f})R(r)=0.
\end{eqnarray}
We define the new variables
\begin{eqnarray}
&& R=\frac{\psi (r)}{r^{\frac{d-1}{2}}},\nonumber\\
&& x=\int \frac{dr}{f(r)},
\end{eqnarray}
and we recast Eq.(\ref{diffe}) into a Schrodinger-like equation of the form
\begin{equation}
\frac{d^2 \psi}{dx^2}+(\omega^2 -V(x))\psi=0.
\end{equation}
Therefore, the expression for the potential is obtained as:
\begin{eqnarray}\label{tt}
V(r)&=&f((d-1) \xi R_{d+1}+\frac{\widetilde{m}^2 (d-1)}{r^2}\nonumber\\
&&+\frac{f^{\prime}}{r}\frac{(d-1)}{2}+\frac{f(d-1)(d-3)}{4 r^2}).
\end{eqnarray}
In Fig.1, we plot the effective potential that the scalar field feels as a function of the distance for different values of coupling constants (Left) and in different dimensions (Right).

\begin{figure*}
\centering
\begin{tabular}{cc}
\rotatebox{0}{
\includegraphics[width=0.5\textwidth,height=0.27\textheight]{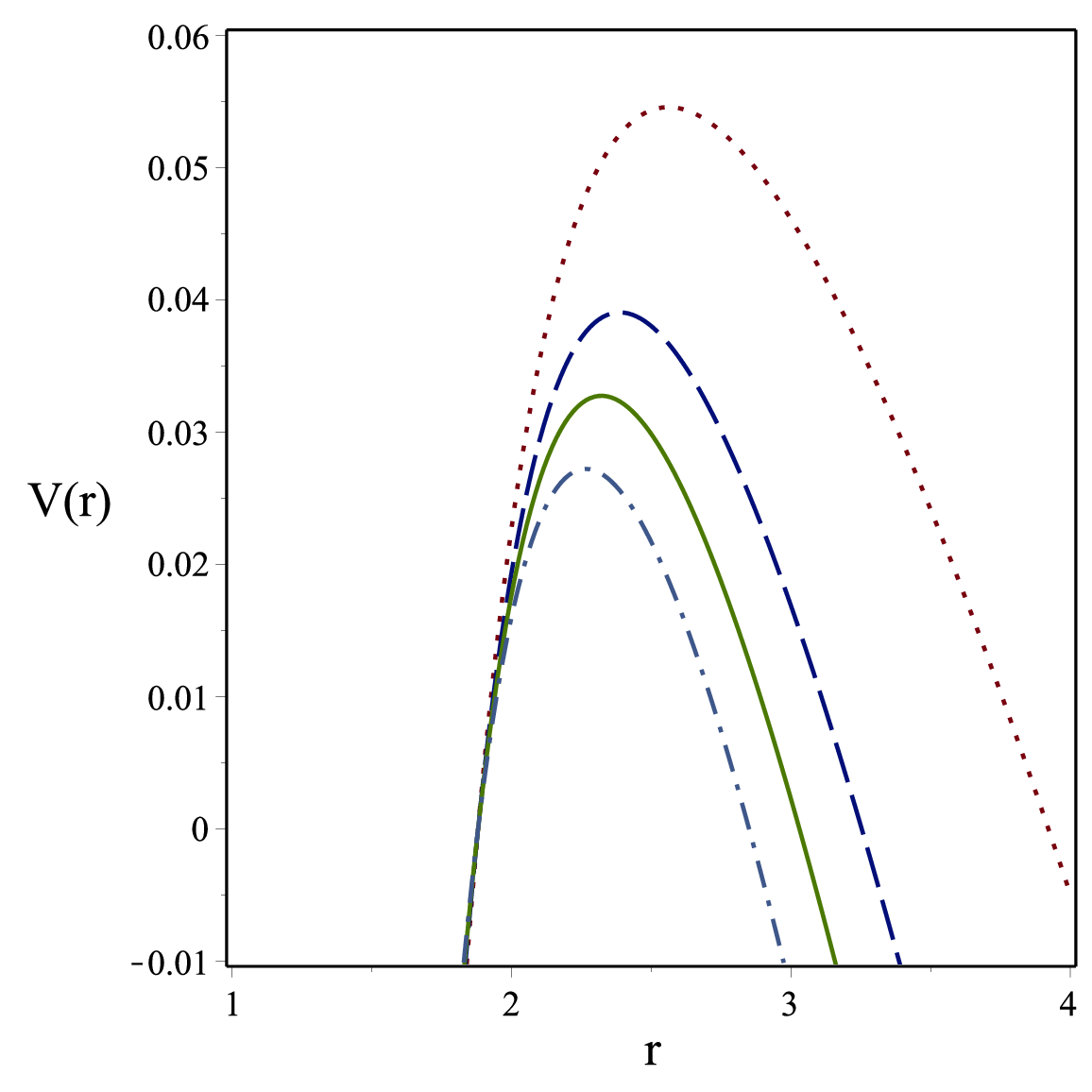}}&
\rotatebox{0}{
\includegraphics[width=0.5\textwidth,height=0.27\textheight]{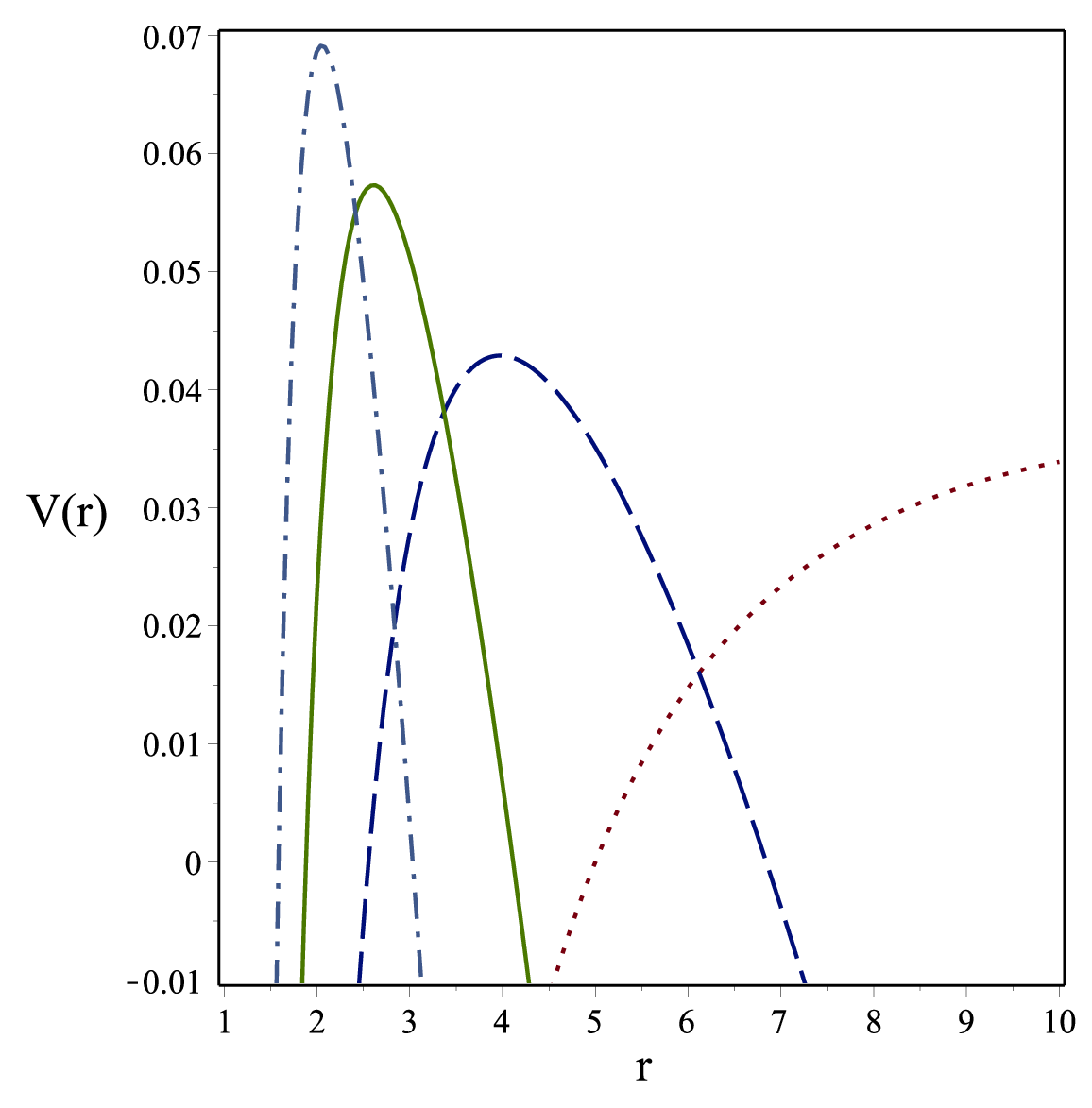}}\\
\end{tabular}

\caption{ Effective potential as a function of distance for different dimensions for $m=1$, $\widetilde{m}=1$, $\xi=0.13$, $l=5$ and for $d=2$ (dot line), $d=3$ (dashed line), $d=4$ (solid line), and $d=5$ (dot-dashed line), (Right hand). Effective potential as a function of distance for different values of coupling constant for $m=1$, $\widetilde{m}=1$, $d=4$, $l=5$ and for $\xi=0.13$ (dotted line), $\xi=0.17$ (dashed line), $\xi=0.19$ (solid line) and $\xi=0.21$ (dot-dashed line), (Left hand).
}\label{figure:RE}
\end{figure*}

As a suggestion for future research, one can find the gray body factor for these black holes and obtain the Hawking temperature by calculating the reflection coefficient and the absorption cross-section for a non-minimally coupled scalar field in the black hole background.

\section{Interior solution}
\label{S3}
After finding the exterior solution for non-rotating black holes with toroidal symmetry in $(d+1)$- dimensions, it may be interesting to ask if we can find any interior solutions for the black holes.
It should be noted that Interior solutions have two different definitions. In the first definition any geodesic may be continued in both directions to an infinite value of the canonical parameter or it ends up to a singular point at a finite value. This solution is also called maximally extended \cite{Berredo}. We use the second definition and assume geometry of our black hole solution describing outside region of a star. The line element for the interior space-time of a static black hole in ($d+1$) dimensions should be written as follows:
\begin{equation}\label{int metric}
d{{s}^{2}}=-{{e}^{2\nu (r)}}d{{t}^{2}}+{{e}^{2\mu (r)}}d{{r}^{2}}+{{r}^{2}}\sum\limits_{i=1}^{d-1}{d\phi _{i}^{2}},
\end{equation}
where, $\nu (r)$ and $\mu (r)$ are two unknown metric functions to be determined.
Here, $\Lambda$ is the negative cosmological constant already introduced in Section III.\\
We consider a static distribution of matter with an energy momentum tensor represented by
\begin{equation}\label{T}
{{T}_{\mu\nu}}=(\rho +P){{U}_{\mu}}{{U}_{\nu}}+P{{g}_{\mu\nu}},
\end{equation}
where, $\rho$ is energy density, ${P}$ is pressure, and ($d+1$)-velocity is given by
\begin{equation}\label{velocity}
{{U}_{\mu}}=-{e^{\nu(r)}}{(1,0,0,....,0)}.
\end{equation}
The explicit form of the energy momentum tensor components is given by
${{T}^{\mu }}_{\nu }=diag(-\rho ,{{P}_{r}},{{P}_{t}},...,{{P}_{t}})$, where ${P}_{r}$ is radial pressure and ${P}_{t}$ is the tangential pressure for the anisotropic fluid (anisotropy was used in the compact star configuration to allow some interesting studies) at the interior of the region.

By solving the Einstein field equations
\begin{equation}\label{Einstein eq}
{{R}_{\mu \nu }}-\frac{1}{2}R{{g}_{\mu \nu }}+\Lambda {{g}_{\mu \nu }}=\pi {{T}_{\mu \nu }},
\end{equation}
in the presence of the negative cosmological constant ($\Lambda<0$) for this anisotropic distribution, we find the following independent equations:
\begin{equation}\label{field1}
\pi \rho +\Lambda =(d-1){{e}^{-2\mu }}(\frac{{{\mu }'}}{r}-\frac{(d-2)}{2{{r}^{2}}}),
\end{equation}
\begin{equation}\label{field2}
\pi {{P}_{r}}-\Lambda =(d-1){{e}^{-2\mu }}(\frac{{{\nu }'}}{r}+\frac{d-2}{2{{r}^{2}}}),
\end{equation}
\begin{eqnarray}\label{field3}
\pi {{P}_{t}}-\Lambda &=&{{e}^{-2\mu }}({\nu }''+{{{\nu }'}^{2}}-{\nu }'{\mu }'+(d-2)(\frac{{\nu }'-{\mu }'}{r})+\frac{(d-2)(d-3)}{2{{r}^{2}}}),
\end{eqnarray}
as well as the ($d-2$) equations, all of which are the same as Eq.(\ref{field3}).\\

Here, the superscript denotes the derivative with respect to $r$ (${\mu }'=\frac{d\mu}{dr}$).
The energy density is then obtained as
\begin{equation}\label{rho1}
\rho =\frac{1}{\pi }((d-1){{e}^{-2\mu }}(\frac{{{\mu }'}}{r}-\frac{(d-2)}{2{{r}^{2}}})-\Lambda ).
\end{equation}
The mass within a radius $r$ is defined as follows:
\begin{equation}\label{m}
m(r)=\int\limits_{0}^{r}{\rho \ d{{V}_{d}}},
\end{equation}
where,
\begin{equation}\label{volume}
d{{V}_{d}}={{(2\pi )}^{d-1}}{{r}^{d-1}}dr.
\end{equation}
To ensure the regular behavior of the mass and energy density functions (finite and positive) at $r=0$, we assume $2\mu (r)=(d-2)\ln r+A{{r}^{d}}$, with $A$ representing a positive constant. \\
Here, we adopt a linear state equation:
\begin{equation}\label{P1}
{{P}_{r}}={{\alpha }_{1}}\rho +{{\alpha }_{2}},
\end{equation}
where, ${{\alpha }_{1}}$ and ${{\alpha }_{2}}$ are two positive and arbitrary constants.
Therefore, the radial pressure is given by:
\begin{equation}\label{P2}
{{P}_{r}}=\frac{{{\alpha }_{1}}}{\pi }(\frac{Ad(d-1)}{2}{{e}^{-A{{r}^{d}}}}-\Lambda )+{{\alpha }_{2}}.
\end{equation}

By using (\ref{field1}) and (\ref{field3}), the unknown function $\nu (r)$ can be obtained:
\begin{equation}\label{nu}
\nu (r)=\frac{-\Lambda }{Ad(d-1)}{{e}^{A{{r}^{d}}}}({{\alpha }_{1}}+1)+\frac{\pi {{\alpha }_{2}}}{Ad (d-1)}{{e}^{A{{r}^{d}}}}+\frac{{\alpha }_{1}A{{r}^{d}}}{2}-\frac{(d-2)}{2}\ln r+C,
\end{equation}
where, C is an integration constant.\\
The energy density ${\rho}$ and the two pressures $({{P}_r,{P}_t})$ are continuous functions of the radial coordinate $r$, indicating that they are well behaved in the interior region.

What we found is the interior space-time solution including the anisotropic perfect fluid which can be now described by the metric (\ref{int metric}).\\
The exterior solution corresponds to a static black hole with toroidal horizon in ($d+1$)-dimensions is written in the following form
\begin{equation} \label{ext metric}
d{{s}^{2}}=-(-\frac{2 m}{d{{r}^{d-2}}}-\frac{2\Lambda {{r}^{2}}}{d(d-1)}) d{{t}^{2}}+
\frac{d{{r}^{2}}} {(-\frac{2 m}{d{{r}^{d-2}}}-\frac{2\Lambda {{r}^{2}}}{d(d-1)})} +{{r}^{2}}\sum\limits_{i=1}^{d-1}{d\phi _{i}^{2}}.
\end{equation}
Matching our interior metric to the exterior metric at the radius of the star $(r=R)$, we get
\begin{equation}\label{C1 cons}
C=-{{\alpha }_{1}}A{{R}^{d}}-\frac{[2\pi {{\alpha }_{2}}-\Lambda ({{\alpha }_{1}}+1)]}{2Ad(d-1)}{{e}^{2A{{R}^{d}}}}+\frac{1}{2}\ln [\frac{-2 m}{d}-\frac{2\Lambda {{R}^{d}}}{d(d-1)}].
\end{equation}
Let us now explore the physical features of our model. The Einstein field equations must in this case satisfy certain physical requirements \cite{faroogh1}.
For the interior solution, the density $\rho $ and the pressures ${P}_{r}$ and ${P}_{t}$ must be positive and finite everywhere. In other words,
\begin{equation} \label{rho2}
\rho =\frac{1}{\pi }[\frac{Ad(d-1)}{2}{{e}^{-A{{r}^{d}}}}-\Lambda ]>0,
\end{equation}
and
\begin{equation}\label{P3}
{{P}_{r}}=\frac{{{\alpha }_{1}}}{\pi }[\frac{Ad(d-1)}{2}{{e}^{-A{{r}^{d}}}}-\Lambda ]+{{\alpha }_{2}}>0.
\end{equation}
Knowing that the cosmological constant is negative, it is obvious that the above relations satisfy both positivity and finiteness.
Also, the gradients of the energy density $\rho $ and the radial and tangential pressures are negative
\begin{equation}\label{rho prime}
\frac{d\rho }{dr}=\frac{-{{A}^{2}}{{d}^{2}}(d-1)}{2\pi }{{r}^{d-1}}{{e}^{-A{{r}^{d}}}}<0,
\end{equation}
and
\begin{equation}\label{P prime}
\frac{{{dP}_{r}} }{dr}=\frac{-{{\alpha }_{1}}{{A}^{2}}{{d}^{2}}(d-1)}{2\pi }{{r}^{d-1}}{{e}^{-A{{r}^{d}}}}<0.
\end{equation}

It can be easily shown that the radial and tangential pressures are decreasing functions of $r$.
The space-time is assumed not to be possess an event horizon. It is indeed for this reason that we select the exponential coefficients in (\ref{int metric}).
Moreover, the radial pressure ${{P}_{r}}$ vanishes but the tangential pressure ${{P}_{t}}$ may not vanish at the boundary $r=R$.
The boundary radius, $R$, can be obtained by letting ${{P}_{r}}(r=R)=0$ using (\ref{P3}), which gives
\begin{equation}\label{radius1}
R=[\frac{-1}{A} \ {\ln({\frac{2}{A d (d-1)}} (\frac{-\pi \alpha_{2}}{\alpha_{1}}+\Lambda))}]^ {\frac{1}{d}}.
\end{equation}

Thus, we found the interior solution for uncharged black holes with toroidal horizons. In what follows, we generalize this investigation to the charged case.

\section{The interior solution for charged black holes}\label{S4}
Our main purpose in this section is to find an interior solution for the charged black holes with toroidal horizons. The
Einstein-Hilbert action coupled to an nonlinear electromagnetic field is captured by the following expression:
\begin{equation}\label{action2}
I=\frac{1}{2\pi }\int{{{d}^{d+1}}x\sqrt{-g}(R-2\Lambda -{{(-{{F}_{\mu \nu }}{{F}^{\mu \nu }})}^{\frac{d}{2}}})}+{{L}_{matter}}.
\end{equation}
For a charged perfect fluid distribution, the Einstein-Maxwell equations with a non-zero cosmological constant are given by:
\begin{equation}\label{Riemann}
{{R}_{\mu \nu }}-\frac{1}{2}R{{g}_{\mu \nu }}+\Lambda {{g}_{\mu \nu }}=\pi (T_{\mu \nu }^{PF}+T_{\mu \nu }^{EM}).
\end{equation}
The energy momentum tensor components for an electromagnetic field are
\begin{equation}\label{T2}
{{T}_{\mu \nu }^{EM}}=\frac{-1}{4\pi }({{F}_{\mu \rho }}F_{\nu }^{\rho }-\frac{1}{4}{{g}_{\mu \nu }}{{F}_{\rho \sigma }}{{F}^{\rho \sigma }}).
\end{equation}
The field strength tensor, ${{F}_{\mu \nu }}$, is related to the current vector in the following way:
\begin{equation}\label{EM}
{{\nabla }_{\mu }}{{F}^{\mu \nu }}=-4\pi {{J}^{\nu }}.
\end{equation}

$T_{\mu \nu }^{PF}$ is the energy momentum tensor for a perfect fluid, which is similar to what we introduced in the previous section. Here, we use the gauge potential ${{A}_{\mu }}=\varphi (r)\delta _{\mu }^{t}$, where $\varphi (r)$ is the electric potential \cite{faroogh3}.
\begin{equation}\label{current}
{{J}^{\nu }}=\sigma (r){{U}^{\nu }},
\end{equation}
where, ${U^{\nu}}$ is the velocity of the charged fluid and
$\sigma (r)$ is the proper charge density of the distribution.\\
We assume the velocity to be:
\begin{equation}\label{velocity2}
{{U}^{\mu }}=\delta _{t}^{\mu }.
\end{equation}
The electromagnetic field tensor is given by
\begin{equation}\label{EM2}
{{F}_{\mu \nu }}={\varphi' (r)}(\delta _{\mu }^{r}\delta _{\nu }^{t}-\delta _{\mu }^{t}\delta _{\nu }^{r}),
\end{equation}
where, the superscript denotes the derivative with respect to $r$ (${\varphi}'=\frac{d\varphi}{dr}$). The explicit form of the energy momentum tensor components in this case will be:
\begin{equation}\label{T2}
{{T}^{\mu }}_{\nu }={{T}^{\mu }}{{_{\nu }}^{PF}}+{{T}^{\mu }}{{_{\nu }}^{EM}}=diag ({\rho}-\frac{E^{2}}{8\pi},{{P}_{r}}-\frac{{{E}^{2}}}{8\pi },{{P}_{t}}+\frac{{{E}^{2}}}{8\pi },... ,{{P}_{t}}+\frac{{{E}^{2}}}{8\pi }),
\end{equation}
where, $E(r)$ is interpreted as the radial component of the static electric field defined by
\begin{equation}\label{E1}
E(r)={\varphi' (r)}{{e}^{(\mu +\nu )}}.
\end{equation}
Again, we adopt a static and axisymmetric line element for describing the internal geometry of the charged distribution, like what we assumed in (\ref{int metric}).
The Einstein-Maxwell field equations for an anisotropic fluid with $\Lambda<0$ will yield:
\begin{eqnarray}\label{eq21}
\pi \rho +\frac{1}{8}{{E}^{2}}+\Lambda =(d-1){{e}^{-2\mu}}(\frac{{\mu }'}{r}-\frac{(d-2)}{2{{r}^{2}}}),
\end{eqnarray}

\begin{eqnarray}\label{eq22}
\pi {{P}_{r}}-\frac{1}{8}{{E}^{2}}-\Lambda =(d-1){{e}^{-2\mu}}(\frac{{\nu }'}{r}+\frac{(d-2)}{2{{r}^{2}}}),
\end{eqnarray}

\begin{equation}\label{eq23}
\pi {{P}_{t}}+\frac{1}{8}{{E}^{2}}-\Lambda ={{e}^{-2\mu}}({\nu }^{\prime \prime} +{{{\nu }'}^{2}}-{\nu }^{\prime} {\mu }^{\prime}+(d-2)(\frac{{\nu }'-{\mu }'}{r})+\frac{(d-2)(d-3)}{2{{r}^{2}}}),
\end{equation}

\noindent and the $(d-2)$ equations, all of which are the same as (\ref{eq23}).\\

There are various choices for $E(r)$ but the acceptable ones are those that leads to a positive and continuous field strength tensor in the interior region. Thus we choose the following exponential function \cite{faroogh1}:
\begin{equation}\label{E2}
{{E}^{2}}=B{{e}^{-K{{r}^{d}}}}-\Lambda.
\end{equation}
where, $K$ is a constant parameter. This function vanishes at the center; i.e, at $E(r=0)=0$ and then $\Lambda =B$, where $B<0$ is a constant parameter.
As in the previous section, we assume that:
\begin{equation}\label{mu2}
2\mu (r)=(d-2)\ln r+K{{r}^{d}},
\end{equation}
so as to ensure a regular behaviour of the mass function at $r=0$.
The energy density for this matter distribution is then given by
\begin{equation}\label{rho4}
\rho (r)=\frac{1}{\pi }((\frac{d(d-1)K}{2}-\frac{1}{8}B){{e}^{-K{{r}^{d}}}}-\frac{7}{8}\Lambda).
\end{equation}
As before, for the linear case, we have:
\begin{equation}\label{P4}
{{P}_{r}}={{\alpha }_{1}}\rho +{{\alpha }_{2}}.
\end{equation}
The second metric coefficient is then obtained in the following form
\begin{equation}\label{nu2}
\nu (r)=(\frac{{{\alpha }_{1}}K}{2}-\frac{(8{{\alpha }_{1}}+1)B}{8d(d-1)}){{r}^{d}}-\frac{1}{2}{{r}^{d-2}}+(\frac{(7{{\alpha }_{1}}-8)\Lambda +8\pi {{\alpha }_{2}}}{8d(d-1)K}){{e}^{K{{r}^{d}}}}+C,
\end{equation}
where, $C$ is the integration constant.
The exterior solution corresponds to a static, charged, black hole is written in the following form:
\begin{equation}\label{ext metric 2}
ds^{2}=-f^{2}(r) d t^{2}+\frac{d r^{2}}{f^{2}(r)}+{{r}^{2}}\sum\limits_{i=1}^{d-1}{d\phi _{i}^{2}}.
\end{equation}
By varying the action (\ref{action2}) without the matter term of $ {{L}_{matter}} $, with respect to the metric $ {{g}_{{\mu}{\nu}}} $ and the gauge field $ {{A}_{\mu}} $ we have
\begin{eqnarray}\label{variying gmunu}
{{R}_{\mu \nu }}-\frac{1}{2}{{g}_{\mu \nu }}(R-2\Lambda )=-{{(-F^2)}^{\frac{d}{2}-1}}(\frac{1}{2}{{g}_{\mu \nu }}{F^2}-d{{F}_{\mu \rho }}F_{\nu }^{\rho }).
\end{eqnarray}
And
\begin{equation}\label{gauge}
{{\partial }_{\mu }}(\sqrt{-g}{{F}^{\mu \nu }}{{(-F^2)}^{\frac{d}{2}-1}})=0
\end{equation}
The right hand of the equation (\ref{variying gmunu}) is the energy-momentum tensor for electromagnetic field and $F^2\equiv F_{{\mu}{\nu}} F^{{\mu}{\nu}}$.

We assume the gauge potential $ {{A}_{\mu }}=\varphi (r)\delta _{\mu }^{t} $ and simplify Eq. (\ref{gauge}) as follows:

\begin{equation}\label{diffpoten}
r {\varphi}^{{\prime}{\prime}}(r)+{\varphi}^{\prime}(r)=0
\end{equation}
It is easy to show that $\varphi (r)=A\ ln(\frac{r}{B})$ is a solution of Eq. (\ref{diffpoten}), where $A$ and $B$ are the integration constants which are related to the electric charge and cosmological constant. Using Eq. (\ref{variying gmunu}), and $\varphi (r)=Q\ ln(\frac{r}{l})$ we obtain lapse function as follows



\begin{equation}\label{F2}
{{f}^{2}}(r)=\frac{-2\Lambda {{r}^{2}}}{d(d-1)}-\frac{2 m}{d{{r}^{d-2}}}-\frac{{{2}^{\frac{d}{2}-1}}{{Q}^{d}}\ln (\frac{-2\Lambda {{r}^{2}}}{d(d-1)})}{{{r}^{d-2}}}
\end{equation}
where, $m$ and $Q$ are the constants related to mass and charge parameters, respectively, and $l=(\frac{-2{\Lambda}}{d(d-1)})^{-1/2}$.
Here, we match our interior metric to the exterior metric and, as a consequence, at the boundary $r=R$, the continuity of the metric yields the following junction condition:

\begin{eqnarray}\label{C2}
C&=&-(\frac{{{\alpha }_{1}}K}{2}-\frac{B(8{{\alpha }_{1}}+1)}{8d(d-1)}) {{R}^{d}+}\frac{1}{2}{{R}^{d-2}}-(\frac{\Lambda (7{{\alpha }_{1}}-8)+8\pi {{\alpha }_{2}}}{8d(d-1)K}){{e}^{K{{R}^{d}}}}\\\nonumber
&&+\frac{1}{2}\ln [\frac{-2\Lambda {{R}^{2}}}{d(d-1)}-\frac{2 m{{R}^{2-d}}}{d}-{{2}^{{\frac{d}{2}}-1}}{{Q}^{d}}{{R}^{2-d}}\ln (\frac{-2\Lambda {{R}^{2}}}{d(d-1)})].
\end{eqnarray}

To achieve a physically acceptable model, the energy density and pressure should be positive and finite in $r<R$. One can see from (\ref{nu2}) and the definition of $\mu$ that these functions are positive if constant $K$ is positive and, further, that the gradients are negative

\begin{equation}\label{rho prime 2}
\frac{d\rho }{dr}=\frac{-1}{\pi }(\frac{{{K}^{2}}{{d}^{2}}(d-1)}{2}{{r}^{d-1}}{{e}^{-K{{r}^{d}}}})<0,
\end{equation}
and
\begin{equation}\label{P prime 2}
\frac{d{{P}_{r}}}{dr}={{\alpha }_{1}}\frac{d\rho }{dr}<0.
\end{equation}
Now, we find the central energy density as
\begin{equation}\label{rho 0}
{{\rho }_{0}}=\rho (r=0)=\frac{1}{\pi }((\frac{d(d-1)K}{2}-\frac{B}{8}){{e}^{-K{{r}^{d}}}}-\frac{7}{8}\Lambda ).
\end{equation}

The boundary radius, $R$, in this case may be obtained by letting ${{P}_{r}}(r=R)=0$. Using (\ref{rho4}) and (\ref{P4}), the radius then turns out to be:

\begin{equation}\label{radius2}
R={{(\frac{-1}{K}\ln (\frac{7\Lambda {{\alpha }_{1}}-8\pi {{\alpha }_{2}}}{4{{\alpha }_{1}}d(d-1)K-{{\alpha }_{1}}B}))}^{\frac{1}{d}}}.
\end{equation}

It is worth noting that the above quantity is finite and positive in all dimensions.
So, we suggested an interior metric for a charged fluid in ($d+1$) dimensions with the toroidal symmetry.

\section{conclusion}\label{S6}
In this paper, the exterior solutions of non-rotating black holes with toroidal symmetry in the ($d+1$)-dimensional space-time were obtained. We used SPSM to calculate the conserved charge (mass) corresponding to the black hole in $(d+1)$-dimensions. Also, we found the entropies for these black holes by using the Wald formula and we obtained the familiar relation between entropy and the area of the event horizon ($S=\frac{A}{4}$). Some of the thermodynamic quantities of this solution were also calculated. These quantities are useful for investigating the thermodynamic and phase transitions of these black holes. The coordinate-invariant quantities found gives us useful geometric information about these space-times. As another facet of the study, we obtained the wave equation associated with a non-minimally coupled scalar field in the black hole background with arbitrary dimensions. In addition, we found an exact interior solution for these black holes for uncharged and charged cases assuming that the interior region is filled with a perfect fluid. Finally, we showed that the proposed $(d+1)$-dimensional interior solutions for the interior region ensures that the energy density and pressures are positive and finite.\\
We hope to continue this study to investigate the thermodynamic properties and phase structure of suggested solutions in this paper. The solution obtained can also be used in AdS/CFT studies.\\

{\large Acknowledgement}: We would like to express our special thanks to K. Hajian for his lavish and useful comments on the solution phase space method.




\end{document}